\documentclass[aps,twocolumn,nofootinbib,preprintnumbers,prl,10pt]{revtex4-2}
\usepackage[utf8]{inputenc}
\usepackage[svgnames]{xcolor}
\usepackage[most]{tcolorbox}
\usepackage{caption}
\tcbset{colback=PaleGreen!0!white, colframe=NavyBlue, 
	highlight math style= {enhanced, 
		colframe=red,colback=red!10!white,boxsep=0pt}
}
\usepackage{relsize}

\usepackage{amsmath}
\usepackage[bottom]{footmisc}
\usepackage{braket}
\usepackage{graphicx}
\usepackage{mwe}
\usepackage[section]{placeins}
\usepackage{framed}
\usepackage{csquotes}
\usepackage{tikz}
\usetikzlibrary{decorations.markings}
\usetikzlibrary{decorations.pathmorphing}
\definecolor{shadecolor}{rgb}{0.90,0.90,0.90}
\usepackage{hyperref}
\usepackage{bbold}
\usepackage{subcaption}
\usepackage{pifont}
\usepackage{setspace}
\usepackage{amsmath, amssymb, amsthm, float, graphicx,amsfonts}
\usepackage{amsthm}
\usepackage{array, boldline, makecell, booktabs}

\theoremstyle{definition}

\hypersetup{colorlinks=true, linkcolor=DarkRed, citecolor=DarkRed,urlcolor=Indigo,linktocpage}
\def\beq{\begin{eqnarray}}\def\eeq{\end{eqnarray}}
\def\be{\begin{equation}}\def\ee{\end{equation}}
\def\bs{\begin{split}}\def\es{\end{split}}
\def\g{\gamma}

\def\s{\sigma}

\def\a{\alpha}
\def\e{\epsilon}

\def\d{\delta}

\def\D{\Delta}

\def\Dphi{\Delta_\phi}

\usepackage{bm}

\begin{document}

\title{\bf Crossing Symmetric Dispersion Relations for Mellin Amplitudes}
\author{\!\!\!\! Rajesh Gopakumar$^{a}$\footnote{rajesh.gopakumar@icts.res.in},~Aninda Sinha$^{b}$\footnote{asinha@iisc.ac.in} and Ahmadullah Zahed$^{b}$\footnote{ahmadullah@iisc.ac.in}\\ ~~~~\\
\it ${^a}$International Centre for Theoretical Sciences (ICTS-TIFR),\\
\it Shivakote, Hesaraghatta Hobli,
Bangalore North, India 560 089\\
\it ${^b}$Centre for High Energy Physics,
\it Indian Institute of Science,\\ \it C.V. Raman Avenue, Bangalore 560012, India. }
\begin{abstract}{We consider manifestly crossing symmetric dispersion relations for Mellin amplitudes of scalar four point  correlators in conformal field theories (CFTs). This allows us to set up the non-perturbative Polyakov bootstrap for CFTs in Mellin space on a firm foundation, thereby fixing the contact term ambiguities in the crossing symmetric blocks. Our new approach employs certain ``locality" constraints replacing the requirement of crossing symmetry in the usual fixed-$t$ dispersion relation. Using these constraints we show that the sum rules based on the two channel dispersion relations and the present dispersion relations are identical. Our framework allows us to connect with the conceptually rich picture of the Polyakov blocks being Witten diagrams in anti-de Sitter (AdS) space. We also give two sided bounds for Wilson coefficients for effective field theories in AdS space.}
\end{abstract}
\maketitle

{\noindent \bf Introduction}: 
Conformal field theories are central to our modern understanding of strongly interacting systems in high energy physics, condensed matter physics and statistical mechanics. It is therefore crucial to develop calculational tools to extract the dynamics of these theories. 
Four point correlation functions in CFTs are constrained by crossing symmetry and unitarity, in addition to the conformal symmetries. The usual bootstrap consistency conditions impose crossing symmetry on the conformal block expansion and this gives strong constraints on the scaling dimensions of operators as well as their three point functions. 

An alternative approach is to make the crossing symmetry manifest at the outset and instead impose other consistency requirements. 
In 1974, Polyakov \cite{Pol} had proposed a version of the conformal bootstrap which used crossing symmetric blocks, while consistency with the operator product expansion (OPE) led to dynamical constraints. This idea lay dormant until recently \cite{ks,usprl}, when the power of this approach was brought out by working in Mellin space \cite{usprl, usothers}. The latter is the natural habitat for conformal correlators, playing the role that momentum space does for flat space scattering amplitudes \cite{mack, mellinjoao, mellinpau, rastellizhou, alday}. 

An attractive feature of this approach is that the basis of Polyakov is essentially that of exchange Witten diagrams used in the AdS/CFT correspondence  \cite{usprl, usothers}. Nevertheless, the non-perturbative validity of this approach remained obscure. Indeed, in \cite{RGAS}, it was pointed out that contact terms need to be added to the basis; however it was not clear what would fix these terms.  In the special case of 1d CFT, this issue was resolved in \cite{mazacpaulos, mazac, PFKGASAZ}; unfortunately these methods do not carry over to higher dimensions. Recent works \cite{joaopaper, joaopaper2, cmrs, sleight} have endeavoured to clarify the non-perturbative validity of the Polyakov bootstrap in $d\geq 2$ by considering dispersion relations in Mellin space. However, these relations exhibit only two channel symmetry, with crossing symmetry additionally imposed, thereby going against the spirit of Polyakov's original work. A non-perturbative version of the crossing symmetric bootstrap would no doubt shed light on the very successful numerical developments arising from bootstrapping position space correlators \cite{rrtv, prv}. Furthermore, through the connection to Witten diagrams, such an approach is the natural one for CFTs with a large radius AdS dual. 

In this letter, we will consider a manifestly crossing symmetric dispersion relation for Mellin amplitudes, focusing on identical scalars. This builds on certain relatively obscure investigations in the 1970s \cite{AK}, whose utility was recently demonstrated in the context of quantum field theories \cite{ASAZ}. 
In this approach, we will employ a crossing symmetric parametrisation of the Mellin variables $(s,t)$ (which are analogous to the usual Mandelstam invariants) and write the corresponding dispersion relation. This is to be contrasted with the more conventional fixed $t$ dispersion relation, which has symmetry manifest  only in the $(s,u)$ channels \cite{joaopaper, joaopaper2}.  

The price we pay for making crossing symmetry manifest is the potential presence of ``non-local" singularities. More precisely, our amplitude will generally have unphysical poles in the Mellin variables (or rather, crossing symmetric combinations thereof). Thus in our approach, we need to impose ``locality constraints'' eq.\eqref{eq:sumrule_gen_Lmn} for consistency. This then leads to our central claim:



{\emph{Once the locality constraints are imposed, the crossing symmetric dispersion relations admit a Witten diagram expansion and, moreover, all contact terms are fixed, as proposed in \cite{RGAS}, up to a constant.}}

We should emphasise that the way the Witten diagrams emerge from the crossing symmetric kernel is quite nontrivial and only happens after imposition of the constraints. Comparing with the approach of \cite{joaopaper, joaopaper2}, our locality constraints appear to be equivalent to their sum rules arising from requiring crossing symmetry (whereas their expressions do not have the unphysical poles above). This demonstrates that the expansion in Witten diagrams is robust and captures the requirements of both crossing symmetry and locality, as might be expected. In both our approach, as well as that of the fixed $t$ dispersion relations, one then proceeds to impose the so-called Polyakov conditions. This is the requirement that there are no spurious double trace operators in the spectrum. It enables us to bootstrap the dynamical conformal data of scaling dimensions and OPE coefficients. 

We note that, in the context of QFTs, the analogous locality constraints \cite{note} were also argued \cite{ASAZ} to be equivalent to the crossing symmetry conditions for the fixed $t$ dispersion relation for effective field theories \cite{TWZ, caron}.
Analogously we will also consider bounds on the Wilson coefficients of the Mellin amplitudes for effective scalar field theories in AdS space. We give preliminary results for two sided bounds similar to those for QFTs \cite{TWZ, caron, ASAZ}.

{\noindent \bf{Crossing symmetric dispersion relation}}:
We define  the Mellin amplitude $\mathcal{M}(s_1,s_2)$ for a four point identical scalar CFT correlator as
\be\nonumber
\mathcal{G}(u,v)=\int \frac{ds_1}{2\pi i} \frac{ds_2}{2\pi i} u^{s_1+\frac{2\Dphi}{3}}v^{s_2-\frac{\Dphi}{3}}\mu(s_i) \mathcal{M}(s_1,s_2)\,,
\ee
where $s_3=-s_1-s_2$, the measure factor $\mu(s_i)=\Gamma^2 \left(\frac{\Dphi}{3}-s_1\right)
\Gamma^2 \left(\frac{\Dphi}{3}-s_2\right) \Gamma^2 \left(\frac{\Dphi}{3}-s_3\right)$ and $\mathcal{G}(u,v)$ is the position space conformal correlator. $\Delta_\phi$ is the scaling dimension of the external scalars. $\mathcal{M}(s_1,s_2)$ is analogous to the usual flat space scattering amplitude and $s_1,s_2$ are analogous to the Mandelstam invariants (related to the usual ($s,t$) by a shift \cite{mellindef}).  Motivated by  S-matrix amplitudes, where we can write a crossing symmetric dispersion relation \cite{AK, ASAZ}, we will write a similar relation for $\mathcal{M}(s_1,s_2)$, assuming its nonperturbative existence in a region of the complex plane  \cite{joaopaper}. We consider hypersurfaces  
$\left(s_{1}-a\right)\left(s_{2}-a\right)\left(s_{3}-a\right)=-a^3,$ with $a$ being a real parameter. The $s_i$'s can then be parametrized as
\be
\label{eq:skdef}
s_{k}(z,a) =a- \frac{a \left(z-z_{k}\right)^{3}}{z^{3}-1}, \quad k=1,2,3\,,
\ee
where $z_k$ are cube roots of unity. Note that $a=\frac{s_1 s_2 s_3}{s_1s_2+s_2s_3+s_3s_1}$ is crossing symmetric.  $\mathcal{M}(s_1,s_2)$ is an analytic function of $(z,a)$. Unlike the S-Matrix amplitude, CFT Mellin amplitudes have poles in the twist of the exchanged operator (with accumulation points) rather than cuts. 
Instead of a dispersion relation in $s$ for fixed $t$, we now write (a twice subtracted) relation in $z$, for fixed $a$, as in \cite{AK, ASAZ}.
In our case, this becomes, following steps similar to \cite{AK, ASAZ}, see supplementary material A of \cite{ASAZ}
\be
\label{eq:disperCFT}
\begin{split}
\mathcal{M}(s_1, s_2)=\alpha_{0}+\frac{1}{\pi} \int_{\tau ^{(0)}}^{\infty}& \frac{d s_1^{\prime}}{s_1^{\prime}} \mathcal{A}\left(s_1^{\prime} ; s_2^{\prime}\left(s_1^{\prime} ,a\right)\right)\\
&\times H\left(s_1^{\prime} ;s_1, s_2, s_3\right)\,,
\end{split}
\ee
where $\alpha_0=\mathcal{M}(s_1=0, s_2=0)$ is a subtraction constant, $\mathcal{A}\left(s_1; s_2\right)$ is the s-channel discontinuity, and
\be\label{kernel}
\begin{split}
&H\left(s ; s_1, s_2,s_3\right)=\left(\frac{s_1}{s-s_1}+\frac{s_2}{s-s_2}+\frac{s_3}{s-s_3}\right)\\
&s_{2}^{\prime}\left(s, a\right)=-\frac{s}{2}\left(1 - \left(\frac{s+3 a}{s-a}\right)^{1 / 2}\right)\,,
\end{split}
\ee
which defines the crossing symmetric kernel $H$ \cite{AK, ASAZ}. Here we have solved for $s_{2}^{\prime}$ for fixed $a$ in terms of the other independent variable $s_1^{\prime}$ from the defining relation below eq.(\ref{eq:skdef}). Note that eq.(\ref{eq:disperCFT}) is manifestly three channel crossing symmetric since the dependence of $s_2^{\prime}$ is only through $a$. In the supplementary material, we have numerically investigated how well eq. \eqref{eq:disperCFT} represents the $2d$-Ising model Mellin amplitude. We have also examined there, the conformal partial wave expansion from which we can work out the domain of $a$ where \eqref{eq:disperCFT} converges. Let $\tau^{(0)}$ be the starting point of the chain of poles in $s_1$.  Following \cite{joaopaper2}, the conformal partial wave expansion converges when  $-\frac{\tau ^{(0)}}{3}\leq Re(a) <\frac{2 \tau ^{(0)}}{3}$  for $\tau^{(0)}>0$ and $\tau ^{(0)}<Re(a) <\frac{2 \tau ^{(0)}}{3}$ for $\tau^{(0)}<0$. 

{\bf \noindent Locality constraints:}
The Mellin amplitude of identical scalars has complete crossing symmetry and hence admits a manifestly symmetric expansion 
\be
\label{eq:cpqdef}
\mathcal{M}(s_1,s_2) = \sum_{p, q=0}^{\infty} {\mathcal M}_{p, q} x^{p} y^{q} = \sum_{p, q=0}^{\infty} {\mathcal M}_{p, q} x^{p+q}a^q \,,
\ee
with $x=-\left(s_1 s_2 + s_2 s_3+s_3 s_1\right)$, $y=-s_1 s_2 s_3$ (with $y=ax$). 
It is important that only positive powers of $x,y$ appear in this expansion around the crossing symmetric point
$(x=y=0)$ so as not to have unphysical singularities. However, as we will see, this will not be obviously evident for our dispersion relation. Thus we will need to impose the nontrivial constraints,
\be\label{eq:sumrule_gen_Lmn}
\mathcal{M}_{p,q}=0\,, p <0  \,.
\ee
%
(Note $q\geq 0$ is inbuilt in our formalism). In the fixed-$t$ dispersion relation, there are no such negative powers and the corresponding sum rules turn out to follow from requiring crossing invariance \cite{examplenull}. In our approach, on the other hand, the constraints in eq \eqref{eq:sumrule_gen_Lmn} are the only additional ones we need to impose, apart from the Polyakov conditions to be discussed below.


{\bf \noindent Crossing symmetric block expansion:}
The s-channel discontinuity in \eqref{eq:disperCFT} comes from the series of poles in the twist (including accumulation points) 
\be\nonumber
\begin{split}
\mathcal{A}(s_1,s_2)=\pi \sum_{\substack{\D,\ell,k }}^{\infty}c_{\D,\ell}^{(k)}P_{\Delta, \ell}\left(\tau_k, s_2\right) \delta\left(\tau_k-s_1\right)\ .
\end{split}
\ee
Here $c_{\D,\ell}^{(k)}$ is proportional to the square of the OPE coefficient and is explicitly given in the supplementary material. The $P_{\Delta, \ell}(s_1, s_2)$ are the Mack polynomials that are the building blocks of the conformal partial wave expansion and are also given for reference in the supplementary material. We have also defined the twist $\tau_k=\frac{\Delta-\ell}{2}+k-\frac{2\Dphi}{3}$ with integer $k\geq 0$. Using these, the dispersion relation \eqref{eq:disperCFT} reads as (note $\tau_k \geq \tau^{(0)}$),
\be\label{eq:disperCFT_final}
\mathcal{M}(s_1, s_2)=\alpha_{0}+\sum_{\substack{\D,\ell,k }}^{\infty}\frac{c_{\D,\ell}^{(k)}}{\tau_k}\mathcal{Q}_{\ell,k}^{(\D)}(a)H(\tau_k;s_1,s_2,s_3)\,,
\ee
where $s_3=-s_1-s_2$ and
\be\label{Qdef}
\begin{split}
\mathcal{Q}_{\ell,k}^{(\D)}(a)\equiv &P_{\Delta, \ell}\left(\tau_k,s_2^{\prime}\left(\tau_k,a\right)\right)\,.
\end{split}
\ee
The final answer \eqref{eq:disperCFT_final} is fully crossing symmetric. Notice that since $s_2^{\prime}\left(\tau_k,a\right)_{s_1=\tau_k}=s_2$, \eqref{eq:disperCFT_final} gives the correct residues at the poles $\tau_k$ in each channel. In the supplementary material, we perform several checks of the convergence of the representation \eqref{eq:disperCFT_final}.

{\noindent \bf Witten diagram expansion}:
We can now relate the block expansion  in  \eqref{eq:disperCFT_final}, to the Witten diagram expansion. 
Note that the $s$-channel Witten diagram can be written in terms of the meromorphic pieces  \cite{RGAS}
\be\label{witt-mer}
M_{\Delta, \ell,k}^{(s)}(s_1, s_2)=P_{\Delta, \ell}(s_1, s_2)\left(\frac{1}{{\tau_k-s_1}}-\frac{1}{{\tau_k}}
\right)\, .
\ee
Here, for convenience, we have subtracted an additional polynomial piece $\propto 1/\tau_k$,  compared to  \cite{RGAS}.
The $t,u$-channel Witten diagrams are related via
$M_{\Delta, \ell,k}^{(t)}(s_1, s_2)=M_{\Delta, \ell,k}^{(s)}(s_2, s_3)$,
$M_{\Delta, \ell,k}^{(u)}(s_1, s_2)=M_{\Delta, \ell,k}^{(s)}(s_3, s_1)$.

The key observation is the following. 
Since $(\frac{1}{\tau_k-s_1}-\frac{1}{\tau_k}) = \frac{1}{\tau_k}\frac{s_1}{\tau-s_1}$, the crossing symmetric kernel $H(\tau_k;s_1,s_2,s_3)$ in \eqref{kernel} also has the same meromorphic pieces, in each channel, as the Witten exchange diagrams. However, the difference between the expansion \eqref{eq:disperCFT_final} and that in terms of the individual Witten diagrams arises in the prefactor. In the Witten diagram \eqref{witt-mer} this is the usual polynomial dependence on the Mellin variables in $P_{\Delta, \ell}(s_1, s_2)$ while that in 
eq. \eqref{eq:disperCFT_final} is the same Mack Polynomial but with an argument $s_2^{\prime}$ as shown in eq. \eqref{Qdef}. 

Now, from eq. \eqref{kernel}, we see that $s_2^{\prime}(\tau_k,a)$ has singularities when $a=\frac{y}{x}=\tau_k$.
These are unphysical since they individually give terms in the expansion which should not be present in the full amplitude. Indeed, as we will see explicitly below, they will give terms with negative powers of $x$ and thus an expansion which is not of the form in eq. \eqref{eq:cpqdef}. Note that $H\left(\tau_k ; s_{1}, s_{2}, s_{3}\right)=\frac{x\left(2 \tau_k-3 a\right)}{x a-x \tau_k+(\tau_k)^3}$, therefore we can expand  \eqref{eq:disperCFT_final} around $a=0,x=0$ to get only positive powers in the expansion \eqref{eq:cpqdef}.  Thus what we need to do is to simultaneously expand \eqref{Qdef} in powers of $a$ and set the nett negative powers of $x$ to zero, as per eq.\eqref{eq:sumrule_gen_Lmn}. Since the meromorphic pieces of  eq. \eqref{eq:disperCFT_final} coincide with those of the exchange Witten diagrams, the nett result is to remove the spurious singularities coming from  $s_2^{\prime}(\tau_k,a)$. This leaves a finite number of polynomial contact pieces, which are all now fixed, up to a constant. 

Let us illustrate the procedure. Consider the difference
\be\nonumber
\mathcal{D}_{\ell}=\frac{1}{\tau_k}\mathcal{Q}_{\ell,k}^{(\D)}(a)H(\tau_k;s_1,s_2,s_3)-\sum_{i=s,t,u}M_{\Delta, \ell,k}^{(i)}(s_1, s_2)\,.
\ee
Given that the Mack Polynomial is of order $\ell$ in the variables, $\mathcal{Q}_{\ell,k}^{(\D)}(a)$ and thus the difference $\mathcal{D}_{\ell}$ will have an unphysical pole, of order $\ell/2$, at $a=\tau_k$. Concretely, take the $\ell=2$ block. The Mack polynomial has a finite expansion $P_{\Delta, \ell}(s_1,s_2)=\sum_{m=0,n=0}^{m+n\leq \ell} (s_1)^m (s_2)^n b_{m,n}^{(\ell)}\,,$. Thus in our case, we have 
$
P_{\Delta, \ell=2}(s_1,s_2)=s_2^2 b_{0,2}^{(2)} + s_1^2 b_{2,0}^{(2)} +s_1 s_2 b_{1,1}^{(2)}+s_1b_{1,0}^{(2)}+s_2b_{0,1}^{(2)}+b_{0,0}^{(2)}\,
$
and therefore we find
\be
\begin{split}
\mathcal{D}_{\ell=2}=\frac{x~ b_{0,2}^{(2)}}{a-\tau_k}+\frac{x~ (b_{0,2}^{(2)}+2 b_{2,0}^{(2)})}{\tau_k}
\end{split}
\ee
As expected, there is a single pole at $a=\frac{y}{x} =\tau_k$, which gives rise to negative powers in $x, y$, as well as additional polynomial pieces. In other words, $\mathcal{D}_{\ell=2}$ takes the form
\be
\mathcal{D}_{\ell=2}=\frac{2 x~ b_{2,0}^{(2)}}{\tau_k}-\frac{y~ b_{0,2}^{(2)}}{ \tau_k ^2}-\left(x~b_{0,2}^{(2)}\right)\sum_{n=2}^{\infty}a^n \tau_k ^{-n-1}.
\ee 
The last term will not contribute once the requirement of \eqref{eq:sumrule_gen_Lmn} is imposed (on the full amplitude). The first two terms are polynomials as expected, and given by
\be
M_{\Delta, \ell=2,k}^{(c)}(s_1, s_2)=\frac{2 x~ b_{2,0}^{(2)}}{\tau_k}-\frac{y~ b_{0,2}^{(2)}}{ \tau_k ^2}\,.
\ee
It should be clear that similar arguments hold for all $\ell$ --- the corresponding ${\cal D}_{\ell}$ will have spurious singularities which when expanded in powers of $a$ give negative powers of $x$ except for a finite number of pieces which are polynomials in $x,y$. Once the locality constraints are imposed on the full amplitude, only these polynomial pieces survive. The $\ell=4$ case is also discussed explicitly in the supplementary material. 

We have therefore argued for the Witten diagram expansion of the Mellin amplitude as 
\be\label{eq:MstWitt}
\mathcal{M}(s_1, s_2)=\sum_{\substack{\D,\ell,k }}^{\infty}c_{\D,\ell}^{(k)}\Big[M_{\Delta, \ell,k}^{(c)}(s_1, s_2)+\sum_{\substack{i=\\s,t,u}}M_{\Delta, \ell,k}^{(i)}(s_1, s_2)\Big]\,.
\ee
$M_{\Delta, \ell,k}^{(c)}(s_1, s_2)$ are the polynomial pieces (for each $\ell$) in $\mathcal{D}_{\ell}$ after removing the spurious singularities at $a=\tau_k$. These are precisely the contact terms which were not fixed in \cite{RGAS}. 
Note that for the $\ell=0$ case we have $M_{\Delta, \ell=0,k}^{(c)}(s_1, s_2)$ is a constant, which gives  the subtraction constant $\alpha_0$. In $d=1$, where there are no spins, this is the only contact term and was discussed in detail in \cite{PFKGASAZ}. Our conclusion then is that the contact term ambiguity anticipated in \cite{RGAS} is fixed for any spin and the Witten diagram expansion holds once the locality constraints are satisfied. 
Further, the explicit form of the contact term shows why it did not contribute to the $O(\epsilon^3)$ results in \cite{usprl, RGAS}, since it can been seen to start contributing only at $O(\epsilon^4)$. 
{\bf \noindent Polyakov conditions and sum rules}:
The measure factor $\mu(s_i)$ in the definition of the Mellin amplitude, introduces additional spurious poles in the full amplitude. 
We must therefore impose the conditions that this expansion not have the poles for double trace operators with $\Delta=2\Dphi+\ell+2p$. Thus $\mathcal M(s_1,s_2)$ must vanish at $s_1=\Dphi/3+p$ \cite{usprl, RGAS, joaopaper}, which are the so-called Polyakov conditions.  
From \eqref{eq:disperCFT_final}, we get
\be\label{eq:sp}
\mathfrak{F}_{p}(s_2)\equiv\mathcal{M}(s_1=\frac{\Dphi}{3}+p, s_2)=0\,.
\ee
{\it Derivation of sum rules in \cite{joaopaper2}:} 
We define the combinations for  $p_{i} \in \mathbb{Z}^{\geq 0}$
\be\label{Omdef}
\begin{split}
&\Omega_{p_1,p_2,p_3}(s_2)\equiv-\frac{\mathfrak{F}_{p_1}(s_2)}{(p_1-p_2)(p_1+p_3+s_2+\frac{2\Dphi}{3})}\\
&-\frac{\mathfrak{F}_{p_2}(s_2)}{(p_2-p_1)(p_2+p_3+s_2+\frac{2\Dphi}{3})}\\
&-\frac{\mathfrak{F}_{p_3}(s_2)}{(p_1+p_3+s_2+\frac{2\Dphi}{3})(p_2+p_3+s_2+\frac{2\Dphi}{3})}\,,
\end{split}
\ee
with $\mathfrak{F}_{p}$ defined in \eqref{eq:sp}. We can now compare with the functional  $\omega_{p_1,p_2,p_3}$ in \cite{joaopaper2}. In our notation 
\be\label{eq:smallomega}
\begin{split}
&\omega_{p_1,p_2,p_3}(s_2)=\sum_{\substack{\D,\ell,k }}^{\infty}c^{(k)}_{\D,\ell}P_{\Delta, \ell}\left(\tau_k,s_2\right) \\
&\times \Bigg[
\frac{ 1}{\prod_{i=1}^2\left(\frac{\Dphi}{3}+p_{i}-\tau_k\right)\left(\frac{\Dphi}{3}+p_{3}+{{\tau}}_k+s_2\right)}\\
&-\frac{ 1}{\prod_{i=1}^2\left(\frac{\Dphi}{3}+p_{i}+s_2+\tau_k\right)\left(\frac{\Dphi}{3}+p_{3}-\tau_k\right)}\Bigg]\,.
\end{split}
\ee
If we now Taylor expand $\omega_{p_1,p_2,p_3}(s_2),~\Omega_{p_1,p_2,p_3}(s_2)$ around $s_2=0$ we get, respectively,
\be
\begin{split}
&\Omega_{p_1,p_2,p_3}(s_2)=\sum_{r=0}^{\infty}\left(s_2\right)^r \Omega_{p_1,p_2,p_3}^{(r)}\\
&\omega_{p_1,p_2,p_3}(s_2)=\sum_{r=0}^{\infty}\left(s_2\right)^r \omega_{p_1,p_2,p_3}^{(r)}\,,
\end{split}
\ee
and find that
$
\omega_{p_1,p_2,p_3}^{(i)}=\Omega_{p_1,p_2,p_3}^{(i)}\,,
$
for $i=1,\cdots,5$.
From $i=6$ onwards, there are nontrivial relations.  To give a flavour of these relations we exhibit, for $r=6$ and $r=7$ (where we have put some of the $p_i$ to zero)   
\be
\omega_{p_1,p_2,0}^{(6)}-\Omega_{p_1,p_2,0}^{(6)}\propto \mathcal{ M}_{-1,2}\,,
\ee
and
\be
\omega_{p_1,0,0}^{(7)}-\Omega_{p_1,0,0}^{(7)}\propto  \mathcal{M}_{-1,2}-\frac{\Dphi(\Dphi+3p_1)}{2\Dphi+3p_1} \mathcal{M}_{-2,3}\,.
\ee
The explicit expressions for the $\mathcal{M}_{p,q}$ appearing here are given below in \eqref{Mdef}.
What these relations indicate is that once the locality constraints \eqref{eq:sumrule_gen_Lmn} are imposed, our Polyakov conditions are the same as those in  \cite{joaopaper2}. 
We have checked that similar relations hold for any $(p_1,p_2,p_3)$ for higher $r$. 

{\bf \noindent Positivity conditions for CFTs:}
Positivity conditions on amplitudes are important for putting bounds on the Wilson coefficients of effective field theories (EFT) in a dual AdS spacetime. We obtain them in our formalism  
following \cite{ASAZ}. From \eqref{eq:disperCFT_final} (see supplementary material), we find
\be \label{Mdef}
\mathcal{M}_{n-m, m} = \sum_{\substack{\D,\ell,k }}^{\infty}c_{\D,\ell}^{(k)}\mathcal{B}_{n, m}^{(\D,\ell,k)}\,, n \geq 1\,.
\ee
Here $\mathcal{B}_{n, m}^{(\D,\ell,k)}$ is given by
\be\label{eq:BellnmU}
\mathcal{B}_{n, m}^{(\D,\ell,k)}=\sum_{q=0}^{m}\mathfrak{U}^{\left(\tau_k\right)}_{n,m,q}(-1)^{m+q} P_{\Delta, \ell;q}\left(\tau_k,0\right)
\ee
where $P_{\Delta, \ell;q}\left(\tau_k,0\right)=\partial_{s_2}^q P_{\Delta, \ell}\left(\tau_k,s_2\right)|_{s_2=0}$ and 
\be
\begin{split}
&\mathfrak{U}^{\left(\tau_k\right)}_{n,m,q}=\frac{(n-q-1)! (m+2 n-3q)}{q! (n-m)! (m-q)! \left(\tau_k\right)^{m+2 n-q+1}}\\&\times\,_{4} F_{3}\left[\begin{array}{c}
\frac{q}{2}+\frac{1}{2},\frac{q}{2},q-m,q+1-\frac{2 n+m}{3} \\
q+1,q+1-n,q-\frac{2 n+m}{3}
\end{array} ; 4\right]\,.
\end{split}
\ee
One can check numerically that {(for $\tau_{k=0}\geq 0$)}, $P_{\Delta, \ell;q}\left(\tau_k,0\right)\geq 0$ -- a similar claim was made in \cite{joaopaper}. Also we find that $\mathfrak{U}^{\left(\tau_k\right)}_{n,m,q}\geq 0$. As in \cite{ASAZ}, we can search for $\chi_{n}^{(r, m)}( \tau_k)$ such that
\be\nonumber
\sum_{r=0}^{m} \chi_{n}^{(r, m)}(\tau_k) \mathcal{B}_{n, r}^{(\D,\ell,k)}=\mathfrak{U}_{n, m, m}^{\left(\tau_k\right)}(-1)^{m+q} P_{\Delta, \ell;q}\left(\tau_k,0\right)\geq 0\,.
\ee
Next, define $\mathcal{M}^{(0)}(s_1, s_2)$, the amplitude after subtracting off the twist zero contributions (if any). Let $\tau^{(0)}$ denote the minimum non-zero twist $\tau_{k=0}$.
Using this in eq.(\ref{Mdef}), we can write down positivity constraints on $\mathcal{M}_{n-m, m}^{(0)}$ similar to those in \cite{ASAZ} 
{(with $\tau^{(0)}\geq 0$)}
\be
\mathcal{M}_{n-m, m}^{(0)}+\sum_{r=0}^{m-1} \chi_{n}^{(r, m)}(\tau^{(0)}) \mathcal{M}_{n-r, r}^{(0)} \geq 0\,.
\ee
A recursion relation for $\chi_n$'s was worked out in \cite{ASAZ}. 
For example \cite{footchi} $
\chi_{n}^{(0,1)}(\tau_k)=\frac{2 n+1}{2 \tau_k}$, $ \chi_{n}^{(0,2)}(\tau_k)=\frac{2 n (n+2)+3}{4 (\tau_k)^2}$, $ \chi_{n}^{(1,2)}(\tau_k)=\frac{2 n+1}{2 \tau_k}
\,.$
These imply for instance $\mathcal{M}_{0,1}^{(0)}>-\frac{3}{2\tau^{(0)}}\mathcal{M}_{1,0}^{(0)}.$
\\
\textit{Two-sided bounds:} If we assume that higher spin partial waves are suppressed, then we can argue for the existence of two-sided bounds. We can check numerically that for $\ell<\ell_*$, there always exists some positive $\beta=O(\ell_*)$ such that
\be
\mathcal{B}_{1, 1}^{(\D,\ell,k)}< \beta \mathcal{B}_{1, 0}^{(\D,\ell,k)}\,,\forall ~ \Dphi \in(\frac{\a}{2},\frac{3\a}{4})\,,
\ee
where $\alpha=d/2-1$.
Therefore, we get a two-sided bound valid for $\alpha<2\Dphi<\frac{3}{2}\alpha$
\be
-\frac{3}{2\tau^{(0)}}\mathcal{M}_{1,0}^{(0)}<\mathcal{M}_{0,1}^{(0)}<\beta \mathcal{M}_{1,0}^{(0)}\,.
\ee
To derive a stronger upper bound analogous to the QFT bounds in \cite{caron, TWZ}, we will need to incorporate the locality constraints as well, a detailed study of which we leave for future work.  
In the case of AdS EFT for a scalar with mass $m$, with the AdS radius given by $R$, we show in the supplementary material that for $m R\gg 1$,
\be
\mathcal{M}_{0,1}^{(A d S)}<\left[1+\frac{\alpha}{2(2 \alpha+3)m^2 R^2}\right] \frac{(10 \alpha+11)}{(2 \alpha+1) \d_0} \mathcal{M}_{1,0}^{(A d S)}\,,
\ee
where $\delta_0$ is related to the EFT scale.
Besides the fact that a two-sided bound in AdS space is novel, the derivation using our approach is algebraically simpler compared to the fixed-$t$ dispersion; in the future, it will be desirable to understand the properties of $\mathcal{B}^{(\Delta,\ell,k)}_{m,n}$ better.


{\bf \noindent Discussion}: The new framework we have developed in this paper has allowed us to make contact with Polyakov's original crossing symmetric bootstrap proposal, and has shown the equivalence, in a nontrivial way, with the results of \cite{joaopaper}. Further, this approach clarifies the underlying reason for the existence of Witten diagram representations for general CFT correlators. To this end, it will be good to broaden the preliminary discussion in the supplementary material on the convergence of the Witten diagram expansion and apply it to new settings. It will also be useful to compare the results developed in this paper with the position space results in eg. \cite{bissi, Paulos:2020zxx}. Mellin space expressions, in turn, permit ready comparison with flat space results. Many of the positivity conditions for (combinations of) Gegenbauer polynomials appear to extend to Mack polynomials. Mathematically it will be useful to develop proofs of these relations which will enable a systematic study of corrections to EFTs in AdS space, in the large radius limit. Finally, we hope that this work will give new impetus to numerical bootstrap studies using the constraints and functionals (such as in \eqref{Omdef}) defined here.

\section*{Acknowledgments} 
We thank Parijat Dey, Kausik Ghosh, Apratim Kaviraj and Miguel Paulos for useful discussions.
R. G.'s research is supported by a J. C. Bose fellowship of the DST as well as project RTI4001 of the Department of Atomic Energy, Government of India,.


\newpage
\onecolumngrid
{\begin{center}\bf \Large{Supplementary material}\end{center} }
\section{Conventions}

In the main text, we have introduced $c^{(k)}_{\D,\ell}=C_{\D,\ell}\mathcal{N}_{\D,\ell}\mathcal{R}_{\D,\ell}^{(k)}$, where $C_{\D,\ell}$ is the OPE coefficient squared as defined in \cite{RGAS} and 
\be\nonumber
\begin{split}
\mathcal{N}_{\Delta, \ell}=\frac{2^{\ell}(\Delta+\ell-1) \Gamma^{2}(\Delta+\ell-1) \Gamma(\Delta-h+1)}{\Gamma(\Delta-1) \Gamma^{4}\left(\frac{\Delta+\ell}{2}\right) \Gamma^{2}\left(\Delta_{\phi}-\frac{\Delta-\ell}{2}\right) \Gamma^{2}\left(\Delta_{\phi}-\frac{2 h-\Delta-\ell}{2}\right)}\,,\mathcal{R}_{\D,\ell}^{(k)}=\frac{\Gamma^{2}\left(\frac{\Delta+\ell}{2}+\Delta_{\phi}-h\right)\left(1+\frac{\Delta-\ell}{2}-\Delta_{\phi}\right)_{k}^{2}}{k ! \Gamma(\Delta-h+1+k)}\,.
\end{split}
\ee
A suitable form for the Mack polynomial can be found in \cite{RGAS}
\be
P^{(s)}_{\D-h,\ell}(s,t)=\sum_{m=0}^\ell\sum_{n=0}^{\ell-m}\mu^{(\D,\ell)}_{n,m}~~\left(\frac{\D-\ell}{2}-s\right)_{m}~~\left(\Dphi-t\right)_{n}\,,
\ee
where 
\be\label{Ap:mu}
\begin{split}
&\mu^{(\D,\ell)}_{n,m}
=\frac{2^{-\ell} \ell! (-1)^{m+n} (h+\ell-1)_{-m} \left(\frac{\ell+\Delta }{2}-m\right)_m (\ell+\Delta -1)_{n-\ell} \left(\frac{\Delta -\ell}{2}+n\right)_{\ell-n} \left(\frac{\Delta -\ell}{2}+m+n\right)_{\ell-m-n} \,}{m! n! (\ell-m-n)!}\\
&{}_4F_3\left(-m,-h+\frac{\Delta-\ell }{2}+1,-h+\frac{\Delta -\ell}{2}+1,n+\Delta -1;\frac{\D+\ell}{2}-m,\frac{\D-\ell}{2}+n,-2 h-\ell+\Delta +2;1\right)\,.
\end{split}
\ee
Our definition $P^{(s)}_{\D-h,\ell}(s,t)$ is different from \cite{RGAS} by a shift in $t$, namely $t\to t-\Dphi$. For our purposes it is convenient to introduce 
$$
P_{\D,\ell}(s_1,s_2)=P^{(s)}_{\D-h,\ell}(s_1+\frac{2\Dphi}{3},s_2+\frac{2\Dphi}{3})\,.
$$
\section{2d Ising model}\label{sec:2dising}
In our conventions, the Mellin amplitude for the 2d Ising model is given by ($\Dphi=1/8$)
 \be\label{eq:2disingamp}
\mathcal{M}^{(2d\text{ Ising})}(s_1,s_2)=\frac{\sqrt{\frac{2}{\pi }} \Gamma \left(-2 s_1-\frac{1}{6}\right) \Gamma \left(-2 s_2-\frac{1}{6}\right) \Gamma \left(-2 s_3-\frac{1}{6}\right)}{\Gamma^2 \left(\frac{1}{24}-s_1\right) \Gamma^2\left(\frac{1}{24}-s_2\right) \Gamma^2 \left(\frac{1}{24}-s_3\right)}\,.
\ee
The dispersion relation Eq. (6) in main text gives the following crossing symmetric pole expansion of the amplitude:
\begin{footnotesize}
\be\label{eq:pole_exp}
\begin{split}
&\mathcal{M}^{(2d~Ising)}(s_1,s_2)=\a_0^{(2d~Ising)}+\sum_{k=0}^{\infty}\Bigg[\left(\frac{1}{\frac{k}{2}-s_1-\frac{1}{12}}+\frac{1}{\frac{k}{2}-s_2-\frac{1}{12}}+\frac{1}{\frac{k}{2}-s_3-\frac{1}{12}}-\frac{3}{\frac{k}{2}-\frac{1}{12}}\right)\times\frac{(-1)^{1-k}}{\sqrt{2 \pi } k! \Gamma \left(\frac{1}{8} (1-4 k)\right)^2}\\
&\times\frac{\Gamma \left(\frac{1}{6} \left(-\frac{1}{2} (6 k-1) \left(\sqrt{1-\frac{48 a}{12 a-6 k+1}}-1\right)-1\right)\right) \Gamma \left(\frac{1}{12} \left(12 k+(6 k-1) \left(\sqrt{1-\frac{48 a}{12 a-6 k+1}}-1\right)-4\right)\right)}{\Gamma^2 \left(\frac{1}{24} \left(1-\left(\sqrt{1-\frac{48 a}{12 a-6 k+1}}-1\right) (6 k-1)\right)\right) \Gamma^2 \left(\frac{1}{24} \left(12 k+(6 k-1) \left(\sqrt{1-\frac{48 a}{12 a-6 k+1}}-1\right)-1\right)\right)}\Bigg]\,.
\end{split}
\ee
\end{footnotesize}

How good is this expansion? To answer this,
we truncate the pole sum $\sum_{k=0}^{\infty}\to \sum_{k=0}^{k_{max}}$ and numerically compare with \eqref{eq:2disingamp}, see table \eqref{tab:2disingamp} (in Mathematica we had to use \$MaxExtraPrecision = 1000 command for better numerical precision). We find that $\a_0^{(2d~Ising)}=-1.48589\times 10^{-6}$. As can be seen from the table, the representation is very good for a variety of random values (including complex ones) for $s_1, s_2$.
\\
\textbf{\textit{Important note:}}
Note that there can be a small mismatch very close to the pole position, where the amplitude diverges. This slight mismatch near the poles is due to the truncation of $k$ sum up to $k_{max}$. 
\begin{table}[hbt!]
\begin{center}
\scalebox{0.8}{
\begin{tabular}{|c|c|c|c|c|}
\hline
 $s_1$ & $s_2$ & \text{Exact} & $k_{\max }=100$ & $k_{\max}=400$ \\
\hline
 4.6 & 0.1 & -0.0032343 & -0.0032459 & -0.0032345 \\
\hline
 8.3 & -1.6 & -0.06009 & -0.059382 & -0.060007 \\
\hline
 8.2+2.1 \textit{i} & -1.6-4.3 \textit{i} & 0.04502 -0.02430 \textit{i} & 0.0457-0.0239 \textit{i} & 0.04510 -0.02426 \textit{i} \\
\hline
 3.3+3.1 \textit{i} & 2.3 +6.7 \textit{i} & 0.038004 -0.0405 \textit{i} & 0.038049 -0.039 \textit{i} & 0.038007 -0.0403\textit{i} \\
\hline
 8.2 +9.7 \textit{i} & 16.3 +29.1 \textit{i} & 0.1001-0.111 \textit{i} & 0.1014-0.084 \textit{i} & 0.1005-0.107\textit{i} \\
\hline
3.1 \textit{i} & 0.2 \textit{i} & 0.008639 -0.003339 \textit{i} & 0.0086 -0.00332 \textit{i} & 0.008636 -0.003337\textit{i} \\
\hline
 13+19 \textit{i} & 21+33 \textit{i} & 0.131 -0.144 \textit{i} & 0.134-0.099 \textit{i} & 0.132-0.136\textit{i} \\
\hline
\end{tabular}}
\end{center}
\caption{Numerical comparison between the $2d$ Ising model amplitude and the crossing symmetric pole expansion of the amplitude. For the pole expansion, we have truncated the $k$ sum upto $k_{\max}$ }
\label{tab:2disingamp}
\end{table}
\begin{figure}[hbt!]
  \begin{subfigure}{5cm}
    \centering\includegraphics[width=5cm]{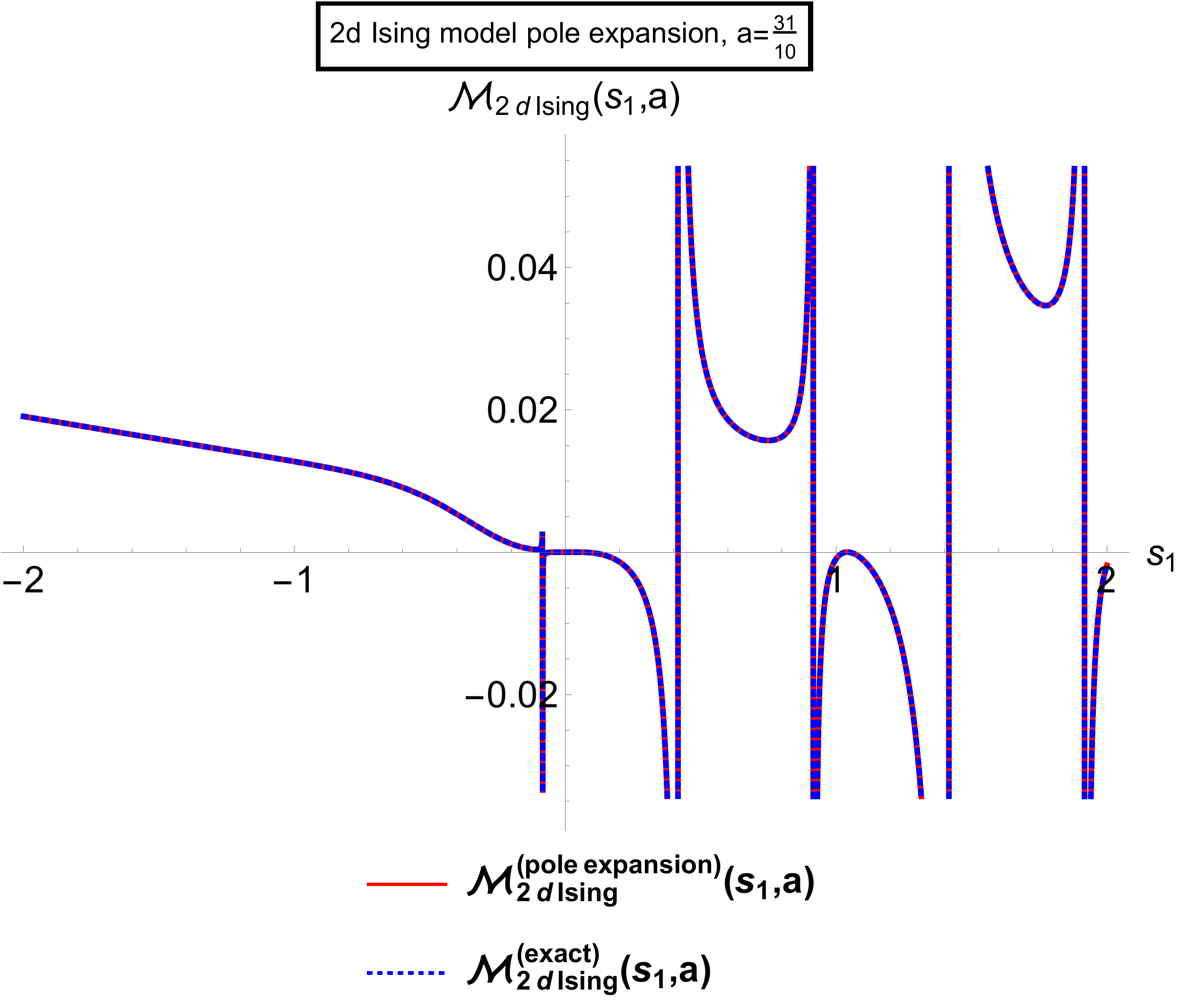}
    \caption{Pole expansion of $2d$ Ising model amplitude for fixed $a=31/10$. k sum truncated at $k_{max}=40$}
  \end{subfigure}
  \begin{subfigure}{5cm}
    \centering\includegraphics[width=5cm]{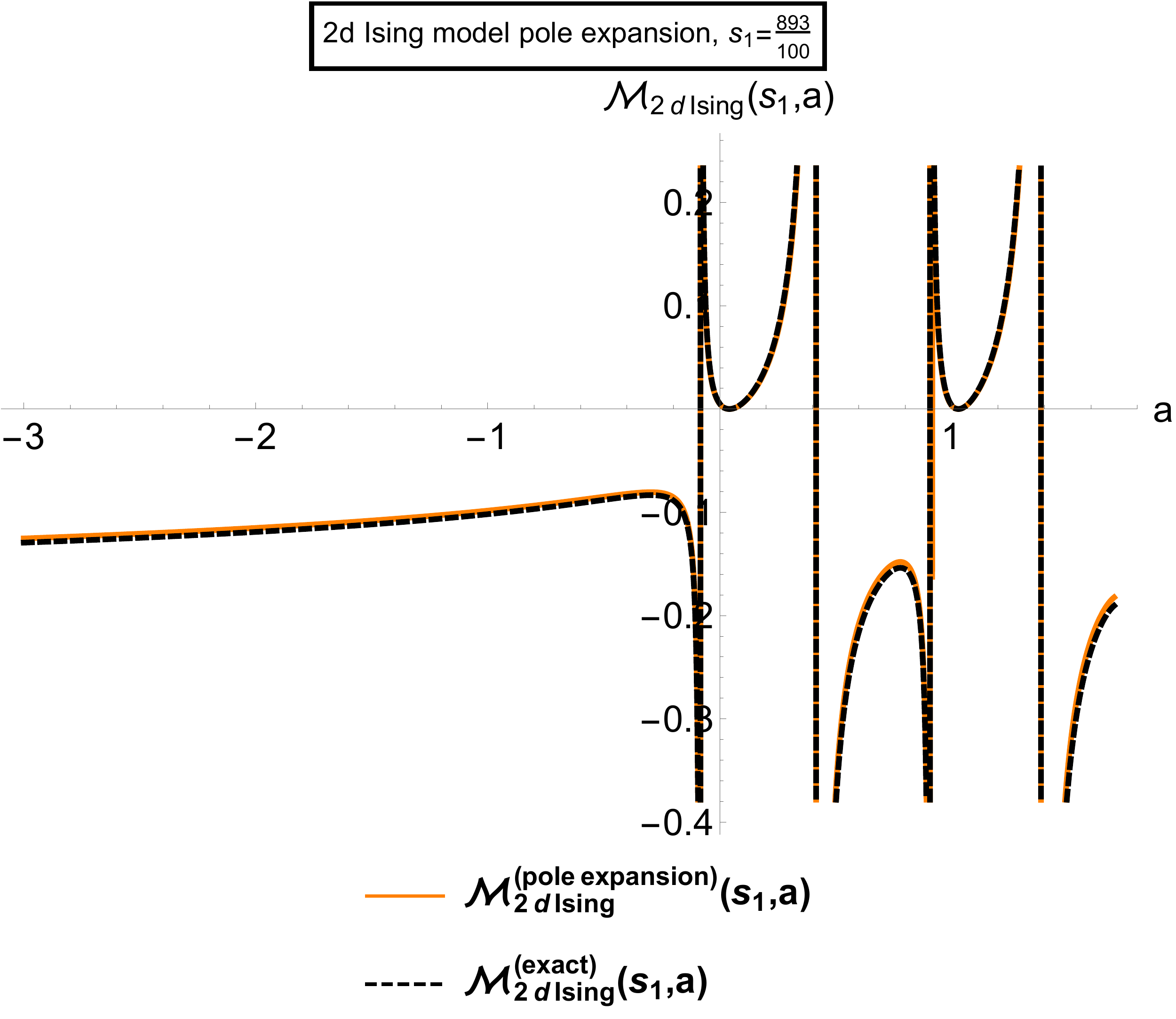}
    \caption{Pole expansion of $2d$ Ising model amplitude for fixed $s_1=893/100$. k sum truncated at $k_{max}=40$}
  \end{subfigure}
  \begin{subfigure}{5cm}
    \centering\includegraphics[width=5cm]{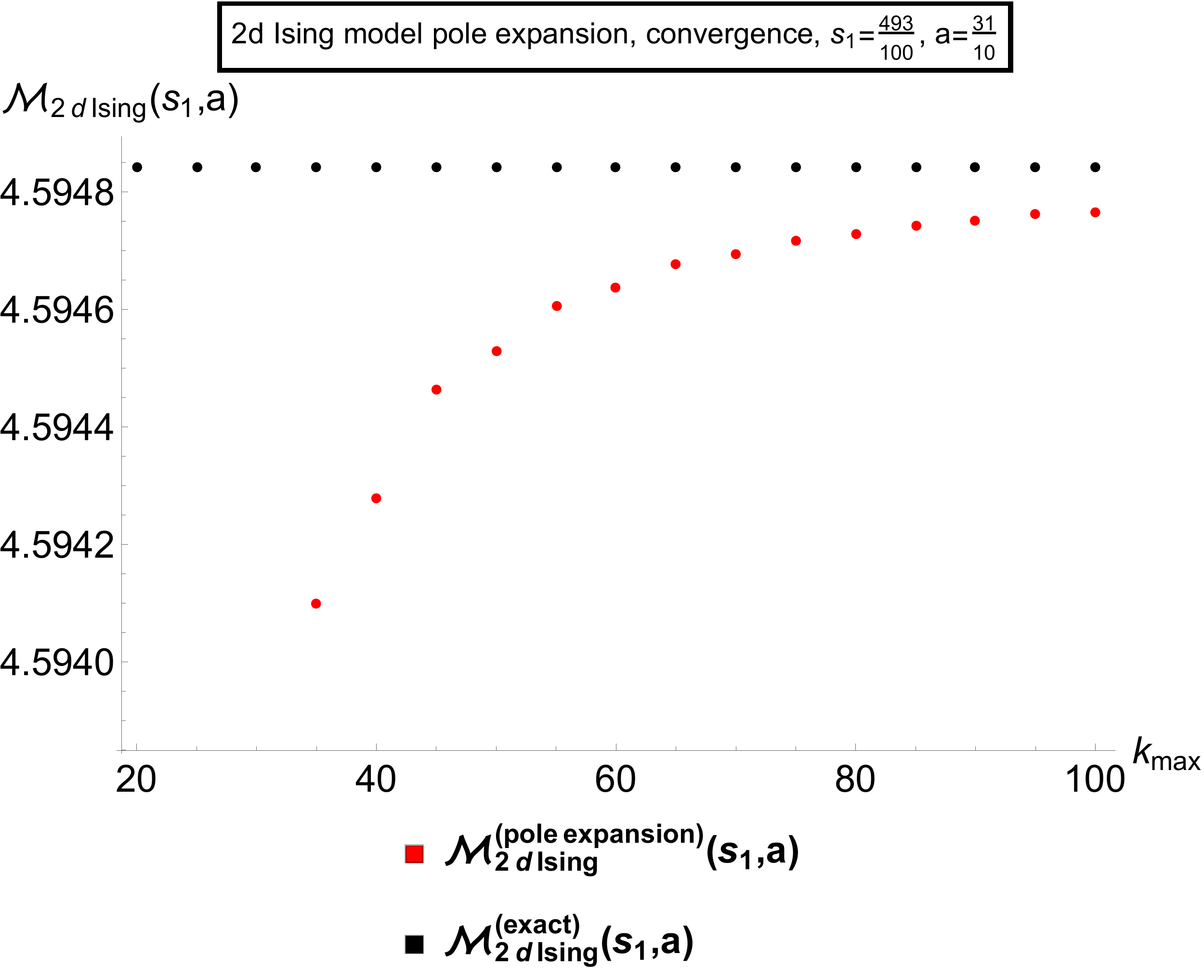}
  \caption{Convergence with $k_{max}$ for $a=31/10,~s_1=493/100$.}
\end{subfigure}
  \caption{Numerical comparisons and convergence of the pole sum (in $k$)}.
  \label{fig:pole_s1a}
\end{figure}
\\
\textbf{\textit{CFT sum rules and 2d Ising model:}}
We illustrate how CFT sum rules, that fix the spectrum, follow directly from the crossing symmetric pole expansion of the above amplitude. Note that we can write \eqref{eq:pole_exp} as
$
\mathcal{M}_{2d~Ising}(a,z)=\sum_{m,n=0}^{\infty}\sum_{k=0}^{\infty}\mathcal{M}_{n-m,m}^{(k)}\,x^n\, a^m\,.
$
The expansion in eq. (4) in main text  tells us that
$
\mathcal{M}_{n-m,m}^{(2d~Ising)}=\sum_{k=0}^{\infty}\mathcal{M}_{n-m,m}^{(k)}=0\,,
$ when $m>n$. For numerical purpose, we truncate the $k$ sum upto $k_{\max}$. As we increase  $k_{\max}$, we get $\mathcal{M}_{n-m,m}^{(2d~Ising)}\to 0$ when $m>n$.
\textbf{\textit{Further numerical checks with 2d Ising model:}} We will put $s_2=s_{2}^{(+)}\left(s_1, a\right)$  in \eqref{eq:2disingamp} and call it $\mathcal{M}_{2d~Ising}^{(exact)}(s_1,a)$. In \eqref{eq:pole_exp}, we will put $s_2=s_{2}^{(+)}\left(s_1, a\right)$ and call it $\mathcal{M}_{2d~Ising}^{(pole~ expansion)}(s_1,a)$, with truncated sum $\sum_{k=0}^{\infty}\to \sum_{k=0}^{k_{max}}$. For $k_{max}=40$ the numerical comparisons are presented in figure \ref{fig:pole_s1a}(a,b). Convergence with increasing $k_{max}$ is presented in figure \ref{fig:pole_s1a}(c).

\section{Convergence of eq (6) and domain of $a$}
We derive the domain of $a$ where eq. (6) in main text converges, following \cite{joaopaper2} (see their Eq. (2.6)). To check convergence, as in \cite{joaopaper2}, we will focus on the large $\ell$ contribution to the sum over spectrum. In this contribution, it is known that operators with dimensions $\D=2\Dphi+\ell+2n+\g(n,\ell)$ contribute with $\g(n,\ell)\sim\ell^{-2\tau^{(0)}-\frac{4\Dphi}{3}}$. 
For $s_1=\frac{\Dphi}{3}+p$ the contribution of these operators to eq. (6) in main text is given by
\be
\sum_{n=0}^{p}\sum_{\ell}\frac{c_{\D,\ell}^{(p-n)}}{\g(n,\ell)}P_{\D,\ell}(\tau_{p-n},s_2'(\tau_{p-n},a))+\left(t,u\right)\text{-channel}\,.
\ee
The $t,u$-channels contribute at subleading order.
From the large $\ell$ behaviour of $
c_{\D,\ell}^{(k)}P_{\D,\ell}(\tau_k,s_2)\sim\ell^{-4\tau_k-\frac{4\Dphi}{3}+2 s_2-1}\,,
$ 
we get the leading term as
\be
\sim \ell^{-2\tau^{(0)}+2 s_2'(\tau^{(0)},a)-1}\,.
\ee
We consider $a$ to be real and readily find that the $\ell$ sum converges if and only if $\tau^{(0)}>s_2'(\tau^{(0)},a)$, which, using the Reduce command in Mathematica, implies the conditions: for $\tau^{(0)}>0$, we have $-\frac{\tau ^{(0)}}{3}\leq a<\frac{2 \tau ^{(0)}}{3}$ and for $\tau^{(0)}<0$, we have $\tau ^{(0)}<a<\frac{2 \tau ^{(0)}}{3}$.
In order to implement the derivative Polyakov condition, one can take derivative of eq. (6) in main text w.r.t $s_1$ and doing a similar analysis for large $\ell$,  one gets
\be
\sim \ell^{\frac{4\Dphi}{3}+2 s_2'(\tau^{(0)},a)-1}+\left(t,u\right)\text{-channel}\,.
\ee
The $\ell$ sum convergence if and only if $s_2'(\tau^{(0)},a)+\frac{2\Dphi}{3}<0$ which leads to the conditions: for $\tau^{(0)}>0$, we have $0<\Delta _{\phi }<\frac{3 \tau ^{(0)}}{4}$ and $ -\frac{\tau ^{(0)}}{3}\leq a<\frac{4 \tau ^{(0)} \Delta _{\phi }^2-6 (\tau ^{(0)})^2 \Delta _{\phi }}{-6 \tau ^{(0)} \Delta _{\phi }+9 (\tau ^{(0)})^2+4 \Delta _{\phi }^2}$, and for $\tau^{(0)}<0$, we have $\tau ^{(0)}<a<\frac{4 \tau ^{(0)} \Delta _{\phi }^2-6 (\tau ^{(0)})^2 \Delta _{\phi }}{-6 \tau ^{(0)} \Delta _{\phi }+9 (\tau ^{(0)})^2+4 \Delta _{\phi }^2}$. {Note that $a$ is always negative in this case.} Away from $s_1=\frac{\Dphi}{3}+p$ we find the leading contribution at large $\ell$ to eq. (6) in main text to be
\be
\sim\frac{\ell^{-4\tau^{(0)}-\frac{4\Dphi}{3}+2 s_2'(\tau^{(0)},a)-1}}{\tau_k-s_1}\,.
\ee
The $\ell$ sum convergence is guaranteed for $2\tau^{(0)}>s_2'(\tau^{(0)},a)$ (we assume $\Dphi>0$): for $\tau^{(0)}>0$, we have $-\frac{\tau^{(0)}}{3}<a<\frac{6\tau^{(0)}}{7}$ and for $\tau^{(0)}<0$, we have $\tau ^{(0)}<a <\frac{6\tau^{(0)}}{7}$.  The bottom line is that we have demonstrated the existence of a range of $a$-values where the conformal partial wave expansion converges. The expansion in eq. (6) in main text and the expansion in terms of Witten diagrams as in eq. (11) in main text differ by the locality constraints which are the same as the conditions arising from imposing crossing symmetry in the fixed-$t$ dispersion relation \cite{joaopaper2}. Assuming the latter set of conditions converge, the Witten diagram expansion eq. (11) in main text can be expected to be convergent.

\section{Convergence of Eq. (6) and Eq. (11): Numerical checks}

\begin{figure}[hbt!]
  \begin{subfigure}{8cm}
    \centering\includegraphics[width=7cm]{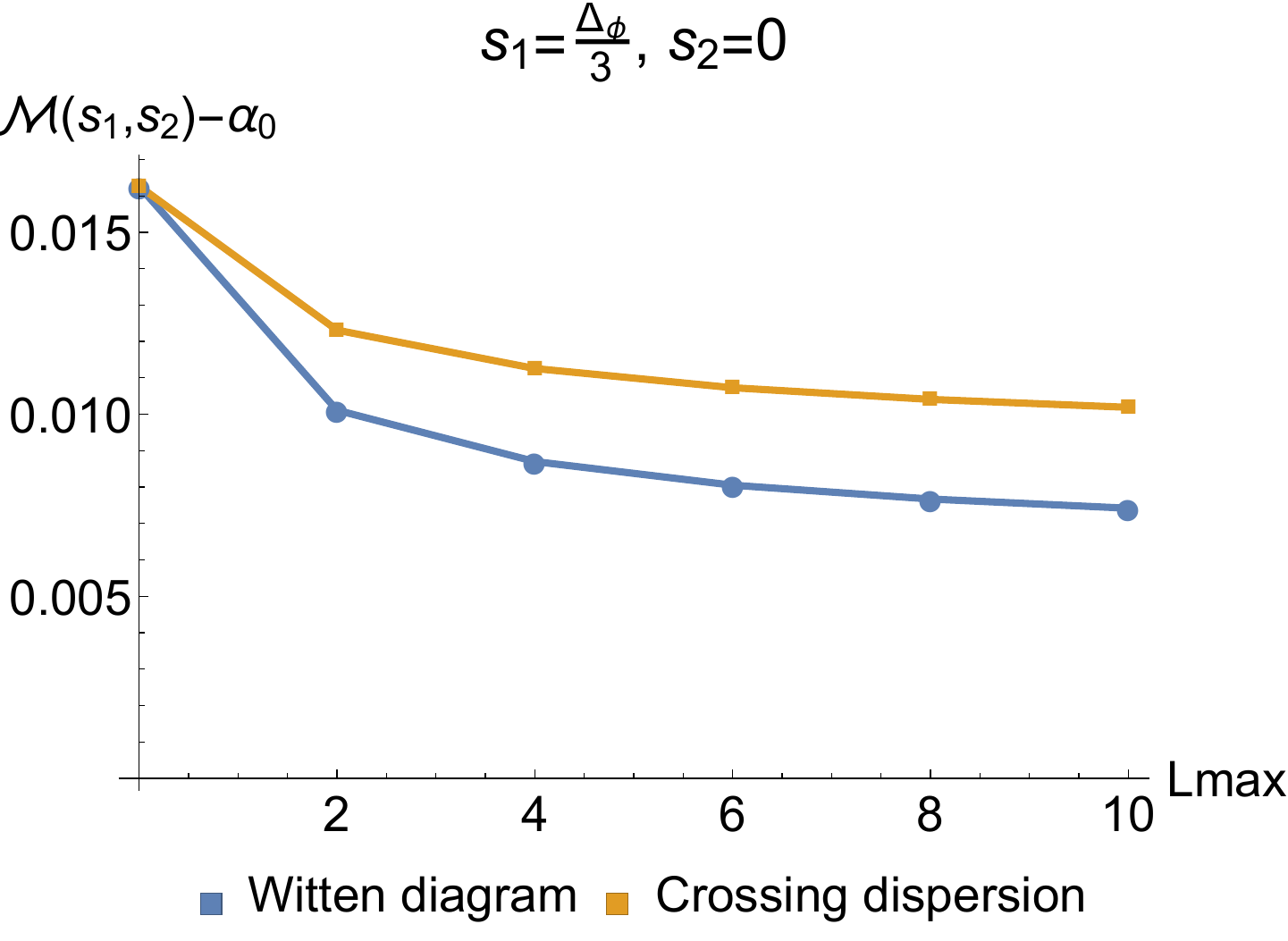}
    \caption{$\mathcal{M}(s_1,s_2)-\a_0$ vs $L_{max}$ for $s_1=\frac{\Dphi}{3}\,,s_2=0$}
  \end{subfigure}
  \begin{subfigure}{8cm}
    \centering\includegraphics[width=7cm]{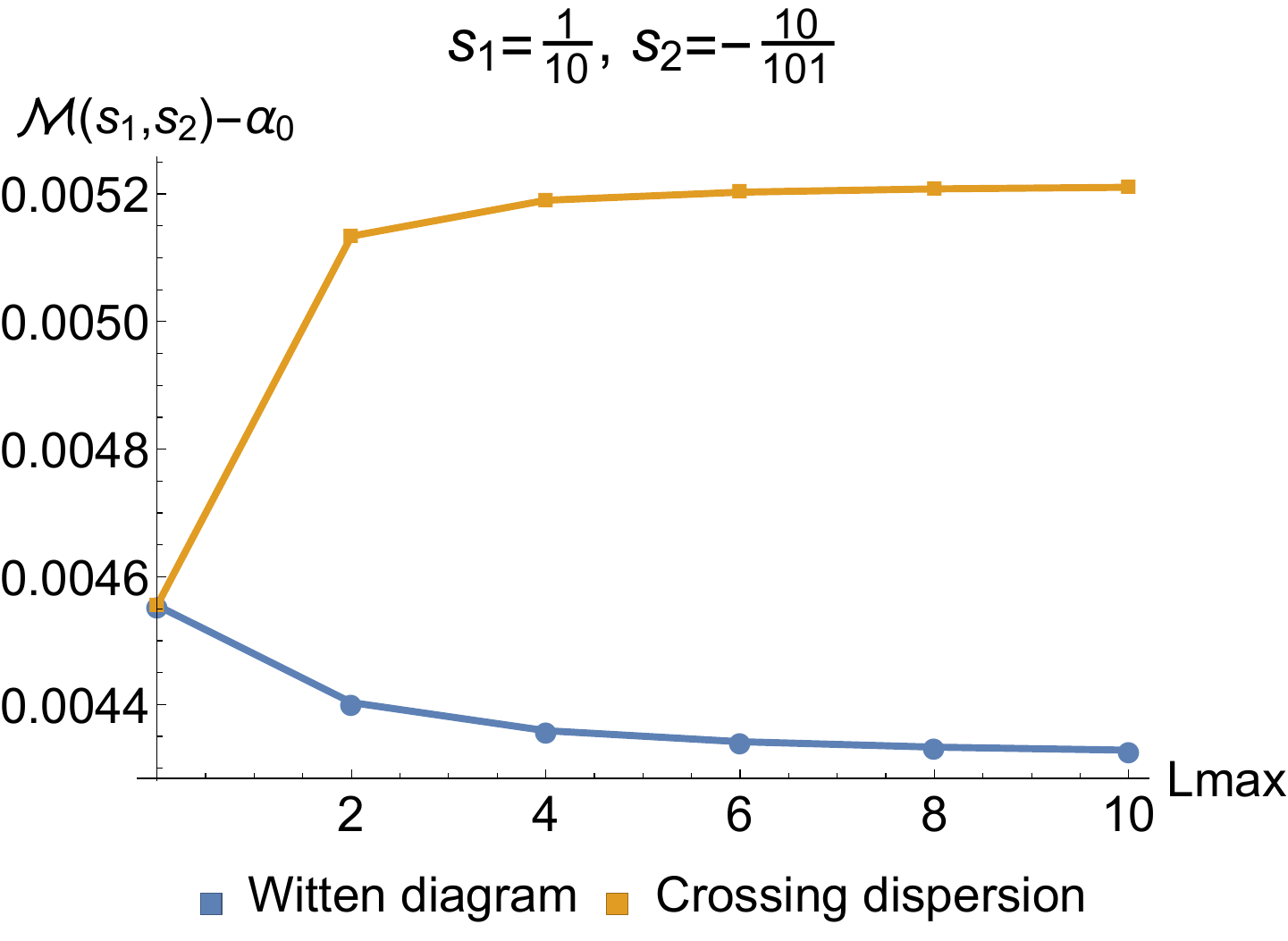}
    \caption{$\mathcal{M}(s_1,s_2)-\a_0$ vs $L_{max}$ for $s_1=\frac{1}{10}\,,s_2=-\frac{10}{101}$}
  \end{subfigure}
  \caption{Convergence of eq eq. (6) in main text and eq. (11) in main text}
  \label{fig:conv5and10}
\end{figure}

Here we will perform only some preliminary numerical checks of convergence of eq. (6) and (11) (in main text) and will find evidence that the sums in both equations are convergent. In order to check convergence, we use the $3d$ Ising model spectrum \cite{duffin} upto spin $10$, namely $\Dphi=0.518149,~\D_{\ell=0}=1.41263, 3.82968, 6.8956, 7.2535, ~\D_{\ell=2}=3., 5.50915, 7.0758,~\D_{\ell=4}=5.02267, 6.42065, 7.38568,~\D_{\ell=6}=7.02849, ~\D_{\ell=8}=9.03102, ~\D_{10}=11.0324$. We truncate the spin sum upto $L_{max}$. The plots with $\mathcal{M}(s_1,s_2)-\a_0$ vs $L_{max}$ is shown in the figure \eqref{fig:conv5and10} and suggests that the sums in both the dispersion relation as well as the Witten diagrams converge. 

We have also checked for convergence using data from the epsilon expansion and again find supporting evidence. For example using $\e=1/10$ (with only double twist spectrum up to $\e^2$), we find $\mathcal{M}\left(s_1=\frac{1}{10},s_2=-\frac{10}{101}\right)-\a_0=0.000310909$ for $L_{max}=2$ and $\mathcal{M}\left(s_1=\frac{1}{10},s_2=-\frac{10}{101}\right)-\a_0=0.000310907$ for $L_{max}=6$ using eq. (11) in main text. 
{If we use eq. (6) in main text, these numbers become $\mathcal{M}\left(s_1=\frac{1}{10},s_2=-\frac{10}{101}\right)-\a_0=0.000310936$ for $L_{max}=2$ and $\mathcal{M}\left(s_1=\frac{1}{10},s_2=-\frac{10}{101}\right)-\a_0=0.000310939$ for $L_{max}=6$. The slight discrepancy between using eq. (6) and (11) (in main text) is due to the fact that the locality constraints (eq. (5) in main text) are not fully satisfied due to considering only the lowest twist operators.}
\section{$\ell=4$ contact term: Explicit expression}\label{ell4cont}
Since $P_{\Delta, \ell=4}(s_1,s_2)=s_1^4 b_{4,0}^{(4)}+s_2^2 \left(s_2^2 b_{0,4}^{(4)}+b_{0,2}^{(4)}\right)+s_1 \left(s_2 \left(s_2 \left(2 s_2 b_{0,4}^{(4)}+b_{1,2}^{(4)}\right)+b_{0,2}^{(4)}\right)+b_{1,0}^{(4)}\right)+s_1^2 \left(s_2 \left(s_2 b_{2,2}^{(4)}+b_{1,2}^{(4)}\right)+b_{2,0}^{(4)}\right)+s_1^3 \left(s_2 \left(b_{2,2}^{(4)}-b_{0,4}^{(4)}\right)+b_{3,0}^{(4)}\right)+b_{0,0}^{(4)}\,,$ we find
\be
\begin{split}
&\mathcal{D}_{\ell=4}=\frac{a^2 x^3 }{y^4 \tau _k \left(a-\tau _k\right)^2}\left(a y^2 b_{0,2}^{(4)} \left(a-\tau _k\right)+a b_{0,4}^{(4)} \left(a^3 x^3+\tau _k \left(y^2 \tau _k \left(7 a-2 \tau _k\right)-a^2 \left(x^3+6 y^2\right)\right)\right)+\right.\\
&(a-\tau _k)\left(y^2 \left(a b_{1,2}^{(4)} \left(3 a-2 \tau _k\right)+2 b_{2,0}^{(4)} \left(a-\tau _k\right)+\left(3 a-2 \tau _k\right) \left(a \tau _k b_{2,2}^{(4)}+b_{3,0}^{(4)} \left(\tau _k-a\right)\right)\right)+\right.\\
&\left.\left.b_{4,0}^{(4)} \left(a-\tau _k\right) \left(2 a^2 x^3+y^2 \tau _k \left(2 \tau _k-3 a\right)\right)\right)\right)\,.
\end{split}
\ee
Throwing away the unphysical terms in $\mathcal{D}_{\ell=4}$, since they cancel once sum over spectrum is performed, one easily obtains $M_{\Delta, \ell=4,k}^{(c)}(s_1, s_2)$:
\be
\begin{split}
M_{\Delta, \ell=4,k}^{(c)}(s_1, s_2)=&\frac{1}{\tau _k^3}\left(-y^2 b_{0,4}^{(4)}-y \tau _k \left(x b_{0,4}^{(4)}+b_{0,2}^{(4)}\right)+2 x \tau _k^4 b_{4,0}^{(4)}+\tau _k^3 \left(2 x b_{3,0}^{(4)}-2 y b_{0,4}^{(4)}+2 y b_{2,2}^{(4)}-3 y b_{4,0}^{(4)}\right)+\right.\\
&\left.\tau _k^2 \left(2 x \left(x b_{4,0}^{(4)}+b_{2,0}^{(4)}\right)+2 y b_{1,2}^{(4)}-3 y b_{3,0}^{(4)}\right)\right)\,.
\end{split}
\ee
We have worked out similar expressions for all even spins up to $\ell=10$.

\section{Derivation of the sum rules of \cite{joaopaper}}

We can Taylor expand Eq.(12) (in main text) equation around $s_2=0$ i.e.
$
\mathfrak{F}_p(s_2)=\sum_{r=0}^{\infty}~s_2^r ~ \mathfrak{F}_p^{(r)}\,.
$
For example 
\be
\begin{split}
\mathfrak{F}_p^{(1)}=&\sum_{\substack{k=0 \\ \D,\ell }}^{\infty} c^{(k)}_{\D,\ell}\Bigg[\frac{2 \left(\Delta _{\phi }+3 p\right)^2 P_{\Delta, \ell;1}\left(\tau_k,0\right)}{9 \tau_k^3-\tau_k \left(\Delta _{\phi }+3 p\right)^2}+P_{\Delta, \ell}\left(\tau_k,0\right) \left(\frac{1}{\tau_k^2}-\frac{9}{\left(\Delta _{\phi}+3 \left(\tau_k+p\right)\right)^2}\right)\Bigg]\,,\\
\end{split}
\ee 
\be
\begin{split}
\mathfrak{F}_p^{(2)}=&\sum_{\substack{k=0 \\ \D,\ell }}^{\infty} c^{(k)}_{\D,\ell}\Bigg[\frac{\left(\Delta _{\phi }+3 p\right){}^2 P_{\Delta, \ell;2}\left(\tau_k,0\right)}{9 \tau_k^3-\tau_k \left(\Delta _{\phi }+3 p\right){}^2}-P_{\Delta, \ell;1}\left(\tau_k,0\right)\left(\frac{9}{\left(\Delta _{\phi }+3 \left(\tau_k+p\right)\right){}^2}-\frac{1}{\tau_k^2}\right)\\
&+P_{\Delta, \ell}\left(\tau_k,0\right) \left(\frac{1}{\left(\frac{\Delta _{\phi }}{3}+\tau_k+p\right){}^3}+\frac{1}{\tau_k^3}\right)\Bigg]\,.
\end{split}
\ee 
We will derive the sum  rules presented in  \cite{joaopaper} by imposing the Polyakov conditions on eq. (6) in main text. We first write down their equation \cite[eq (6.4) or eq (139) in arxiv version]{joaopaper}
\be\nonumber
\begin{split}
F({\g_{13}})=&2 G\left(\left(\gamma _{13}-\frac{\Delta _{\phi }}{3}\right){}^2\right)+\sum_{\substack{k=0 \\ \D,\ell }}^{\infty} c^{(k)}_{\D,\ell}P_{\Delta, \ell}\left(\tau_k, \frac{\Dphi}{3}-\g_{12}\right)\\
&\times\left(\frac{3}{3 \gamma _{13}+2 \Delta _{\phi }-3 \tau_k}-\frac{1}{\gamma _{13}+\Delta _{\phi }-\tau_k}+\frac{3}{3 \tau_k-4 \Delta _{\phi }}+\frac{3}{5 \Delta _{\phi }-3 \tau_k}\right)\,.
\end{split}
\ee
We taylor expand $F(\g_{13})$ around $\g_{13}=\Dphi/3\,,$ (i.e. $s_2=2\Dphi/3$ in our notation)
$
F(\g_{13})=\sum_{r=0}^{\infty}\left(\gamma _{13}-\frac{\Delta _{\phi }}{3}\right)^r F^{(r)}\,.
$
We find that $F^{(r)},~\mathfrak{F}_0^{(r)}, ~\mathcal{M}_{n-m,m}$ are related. For example
\be\nonumber
\begin{split}
&F^{(1)}=\mathfrak{F}^{(1)}_0,~F^{(3)}=\mathfrak{F}^{(3)}_0, ~F^{(5)}-\mathfrak{F}^{(5)}_0=2\mathcal{M}_{-2,3},~F^{(7)}-\mathfrak{F}^{(7)}_0=4 \mathcal{M}_{-4,5}- \mathcal{M}_{-1,3}-\frac{27}{\Dphi^2}\mathcal{M}_{-2,3}+\frac{27}{\Dphi^3}\mathcal{M}_{-1,2}\,,\\
&F^{(9)}
-\mathfrak{F}^{(9)}_0=6\mathcal{M}_{-6,7}-6 \mathcal{M}_{-3,5}-\frac{90}{\Dphi^2} \mathcal{M}_{-4,5}+\frac{9}{\Dphi^2}\mathcal{M}_{-1,3}+\frac{162}{\Dphi^3}\mathcal{M}_{-3,4}+\frac{81}{\Dphi^4}\mathcal{M}_{-2,3}-\frac{243}{\Dphi^5}\mathcal{M}_{-1,2}\,.
\end{split}
\ee 
Similar relations hold for any odd $r$ (we have compared only odd $r$ since for even $r$, $F^{(r)}$ will contain $G^{(r)}(0)$ which is known in terms of the full amplitude, but not apriori (see comments near \cite[eq (139)]{joaopaper}).

\section{Epsilon expansion of $\mathcal{M}_{n-m,m}\,,m>n$}

We will study the epsilon expansion of the locality constraints illustrating with the case of $\mathcal{M}_{-2,3}$. The operator dimensions $\Delta _{\ell ,q}$ 
of the operators ${\cal O}_{\ell, q}$  (with twist $2+q$; $q\in \mathbb{Z}^{\geq 0}$) have an $\e$ expansion 
$
\Delta _{\ell ,q}=2-\epsilon+\ell+2q+\delta _1(q,\ell ) \epsilon+\delta _2(q,\ell )\epsilon ^2 +\dots
$
 and $\Dphi$ has an expansion
$
\Dphi=1-\frac{1}{2}\e+\d_{\phi}^{(2)}\e^2\,.
$
\\
\textbf{\emph{$\mathbf{O(\e^2)}$:}} The OPE coefficients of $\D_{\ell,0}$ starts at $O(\e^0)$;
$
C_{\D_{\ell,0},\ell}=C_{0,\ell}^{(0)}+C_{0,\ell}^{(1)}\e +C_{0,\ell}^{(2)}\e^2+\dots\,.
$ Then $\mathcal{M}_{-2,3}$ starts at $O(\e^2)$. The only operators that contribute are $\D_{\ell,0}$ (higher twists start at $O(\e^4)$, see below). At $O(\e^2)$, we get contributions 
$
A_{2}(0,\ell) C_{0,\ell}^{(0)}\left(\d_1(0,\ell)\right)^2\e^2\,,
$
with $A_{2}(0,\ell)$ a known quantity (the exact form is cumbersome and is not needed). For example 
$
A_2(0,2)=-1825.03
$
and so on. The important point is that $A_{2}(0,\ell)<0,\forall\ell$. In order to satisfy the constraint
$$
\mathcal{M}_{-2,3}=\sum_{\ell=0}^{\infty}A_{2}(0,\ell) C_{0,\ell}^{(0)}\left(\d_1(0,\ell)\right)^2\e^2+O(\e^3)=0
$$
condition at $O(\e^2)$, we must have,
\be
\d_1(0,\ell)=0.
\ee
In other words, the anomalous dimension of twist two operators should start at $O(\e^2)$.
Once the constraint is satisfied at $O(\e^2)$ it is also automatically satisfied at $O(\e^3)$, since the next nonzero contribution to $\mathcal{M}_{-2,3}$ begins at $O(\e^4)$ when $\d_1(0,\ell)=0$.
\\
\textbf{\emph{$\mathbf{O(\e^4)}$:}} At $O(\e^4)$, all operators start contributing. The point to emphasize is that the OPE of operators with $q\neq0$ starts at $O(\e^2)$;
$
C_{\D_{\ell,q},\ell}=C_{q,\ell}^{(2)} \e^2+\dots, (~~q\neq0).
$
It is therefore impractical to solve the constraints at $O(\e^4)$. At this order, the contributions from the twist two operators ($q=0$) is 
$
A_{2}(0,\ell) C_{0,\ell}^{(0)}\left(-2\d_{\phi}^{(2)}+\d_2(q,\ell)\right)^2\e^4
$
and those  from twist four onwards ($q\neq0$) is
$
A_{4}(q,\ell) C_{q,\ell}^{(2)}\left(\d_1(q,\ell)\right)^2\e^4
$
with $A_{4}(q,\ell)$ a known quantity (for example
$
A_{4}(1,2)=-10.22
$
and so on). Some observations are as follows:
$
A_{4}(1,2)<0,~A_{4}(1,4)<0,~
A_{4}(1,\ell)>0 ~\text{for} ~\ell\geq6,~\text{and}~A_{4}(2,2)<0,~
A_{4}(2,\ell)>0 ~\text{for} ~\ell\geq4\dots\,.
$
Therefore  at $O(\e^4)$ 
$$\mathcal{M}_{-2,3}=\sum_{\ell=0}^{\infty}\Bigg[A_{2}(0,\ell) C_{0,\ell}^{(0)}\left(-2\d_{\phi}^{(2)}+\d_2(q,\ell)\right)^2+\sum_{q=1}^{\infty}A_{4}(q,\ell) C_{q,\ell}^{(2)}\left(\d_1(q,\ell)\right)^2\Bigg]\e^4+O(\e^5)\,.$$
Thus the contributions from the leading twist operators, gets cancelled by the higher spin contributions of the higher twist operators. 

\section{Derivation of eq (18)-(19)}

We will follow  the steps in \cite{ASAZ}, but now for our CFT amplitude. We need
\be
H\left(\tau_k ; s_{1}, s_{2}, s_{3}\right)=\frac{27 a^{2} z^{3}\left(3 a-2 \tau_k\right)}{-27 a^{3} z^{3}+27 a^{2} z^{3}\tau_k+\left(z^{3}-1\right)^{2}\left(\tau_k\right)^{3}}\,,
\ee
where $\frac{z^3}{(1-z^3)^2}=-\frac{x}{27 a^2}$. We can now expand in power of $x$, and collect the powers of $a$ to find, very similar to \cite{ASAZ},
\be
\begin{split}
&\mathcal{M}_{n-m, m} \equiv \sum_{\substack{\D,\ell,k }}^{\infty}c_{\D,\ell}^{(k)}\mathcal{B}_{n, m}^{(\D,\ell,k)}\,, n \geq 1\,.\\
&\mathcal{B}_{n, m}^{(\D,\ell,k)}=\sum_{j=0}^{\ell/2} \frac{1}{\tau_k^{2 n+m+1}} \frac{p_{\ell}^{(j)}\left(\xi_{0}\right)}{j !}(4\xi_{0})^j\times \frac{(3 j-m-2 n)(-n)_{m}}{(m-j) !(-n)_{j+1}} \, .
\end{split}
\ee
Here $p_{\ell}^{(j)}\left(\xi_{0}\right)=\left.\frac{\partial^{j} P_{\D,\ell}\left(\tau_k,\frac{1}{2} \left(\sqrt{\xi }-1\right) \tau _k\right)}{\partial \xi^{j}}\right|_{\xi=\xi_{0}}\, $ and we have replaced $s_2'(\tau_k,a)$ by solving $\sqrt{\xi}=1+\frac{2~s_2'(\tau_k,a)}{\tau_k}$ and $\xi_{0}=1$. For example $p_{\ell}^{(1)}\left(\xi_{0}\right)=\frac{\tau _k P_{\D,\ell;1}\left(\tau_k,0\right)}{4}\,,p_{\ell}^{(2)}\left(\xi_{0}\right)=\frac{\tau_k\left(\tau_k P_{\D,\ell;2}\left(\tau_k,0\right)-2 P_{\D,\ell;1}\left(\tau_k,0\right)\right)}{16}$, and more generally \cite{ASAZ}, we find equation eq (19) in main text. Notice that for $m>n$, we have $\ell\geq 2n$, which follows on writing $(-n)_m/(-n)_{j+1}=\Gamma(-n+m)/\Gamma(-n+j+1)$ so that for this to not vanish we need $j\geq n$. Since the argument of the degree-$\ell$ polynomial is $\sqrt{\xi}$, we will need $\ell\geq 2n$ as noted in \cite{AK, ASAZ}.

\section{Two sided bounds, EFT in AdS}
In the large $\nu=\D-h,~ s_1$ limit ($s_1=\d+4\Dphi/3=s-2\Dphi/3$), we have 
\be
\begin{split}
P_{\nu, \ell}^{(s)}(s,t)=&\frac{8^{-\ell} \ell ! s^{\ell}}{(h-1)_{\ell}}\Bigg[C_{\ell}^{(h-1)}(x)-\frac{(h-1)}{\nu^{2}} C_{\ell-2}^{(h)}(x)+O\left(\frac{1}{\nu^4}\right)\Bigg]+O\left(s^{\ell-1}\right)\,.
\end{split}
\ee
We assume $s\gg \nu^2$. Following \cite{smat3, PHAS} we get the $s$-channel discontinuity 
\be
\mathcal{A}^{(AdS)}(s, t) \approx \sum_{\ell} a_\ell(s) \left( C_{\ell}^{(h-1)}(x)-\frac{(h-1)}{\nu^{2}} C_{\ell-2}^{(h)}(x)\right)\,,
\ee
where
\be
a_\ell(s)=\pi \sum_{\D} C_{\D, \ell} \mathcal{N}_{\D, \ell} \frac{\Gamma\left(2 \Delta_{\phi}+\ell-h\right)}{2 \Delta_{\phi}+\ell} \frac{\sin ^{2} \pi\left[\Dphi-s\right]}{\sin ^{2} \pi\left[\Delta_{\phi}-\frac{\D}{2}\right]}\frac{8^{-\ell} \ell ! s^{\ell}}{(h-1)_{\ell}}\delta\left(s-\frac{\Delta-\ell}{2}-q_{\star}\right)\,.
\ee
where $q_{\star}=\frac{\left(\frac{\D-\ell}{2}-\Delta_{\phi}\right)^{2}}{\ell+2 \Delta_{\phi}}\,.$ Assuming the lower limit of the $\D$ sum is very large ($\D-\ell\sim 2\Dphi$ is the lower limit), the corresponding $\nu$ is given by $\nu_0\approx 2mR$ since $m^{2} R^{2}=\Delta_{\phi}\left(\Delta_{\phi}-2h\right)$. Therefore for large $R$ (upto $\frac{1}{R^2}$), we can write
\be
\mathcal{A}^{(AdS)}(s, t) \approx \sum_{\ell} a_\ell(s) \left( C_{\ell}^{(h-1)}(x)-\frac{(h-1)}{\nu_0^{2}} C_{\ell-2}^{(h)}(x)\right)\,.
\ee
In the limit $s\gg \nu^2$, other corrections will be sub-leading in $s,R$. We can now do a similar analysis as in \cite{ASAZ}.  We note that in that limit
$
\mathcal{B}_{n, 0}^{(AdS,\ell)}=\frac{2}{\pi \delta^{2 n}}\left(1-\frac{\alpha}{\nu_0^{2}}\right)\,,
$
which implies 
\be
\mathcal{B}_{n, 0}^{(A d S,\ell)}=\left(1-\frac{\alpha}{\nu_0^{2}}\right) \mathcal{B}_{n, 0}^{(Flat,\ell)} \text{ or } \mathcal{M}_{n, 0}^{(A d S,\ell)}=\left(1-\frac{\alpha}{\nu_0^{2}}\right) \mathcal{M}_{n, 0}^{(Flat,\ell)}\,.
\ee
Since $\delta_0\gg 1$, we have
$\mathcal{B}_{1,1}^{(A d S,\ell)}=\frac{4 \ell(\ell+2 \alpha)-3(2 \alpha+1)}{\pi(2 \alpha+1) \delta^{3}}-\frac{\alpha[4 \ell(\ell+2 \alpha+2)-3(2 \alpha+2+1)]}{\pi(2 \alpha+2+1) \delta^{3} \nu_0^{2}}\,,$ which gives
\be
\begin{split}
\mathcal{B}_{1,1}^{(A d S,\ell)}\leq \left(1-\frac{\alpha(2 \alpha+1)}{(2 \alpha+3)\nu_0^2}\right) \mathcal{B}_{1,1}^{(Flat,\ell)} \text{ or } \mathcal{M}_{0,1}^{(A d S,\ell)}\leq \left(1-\frac{\alpha(2 \alpha+1)}{(2 \alpha+3)\nu_0^2}\right) \mathcal{M}_{0,1}^{(Flat,\ell)}  \,.
\end{split}
\ee
From \cite{TWZ, caron}
$
\mathcal{M}_{0,1}^{(Flat)}<\frac{10 \alpha+11}{(2 \alpha+1) \delta_{0}} \mathcal{M}_{1,0}^{(Flat)}\,,
$
we get 
\be
\mathcal{M}_{0,1}^{(A d S)}<\left[1+\frac{2 \alpha}{(2 \alpha+3) \nu_{0}^{2}}\right] \frac{(10 \alpha+11)}{(2 \alpha+1) \delta_{0}} \mathcal{M}_{1,0}^{(A d S)}\text{ or }
\mathcal{M}_{0,1}^{(A d S)}<\left[1+\frac{\alpha}{2(2 \alpha+3)m^2 R^2}\right] \frac{(10 \alpha+11)}{(2 \alpha+1) \d_0} \mathcal{M}_{1,0}^{(A d S)}\,.
\ee


\begin{thebibliography}{99}

\bibitem{Pol}
A.~M.~Polyakov,
``Nonhamiltonian approach to conformal quantum field theory,''
Zh. Eksp. Teor. Fiz. \textbf{66}, 23-42 (1974)

\bibitem{ks}
K.~Sen and A.~Sinha,
``On critical exponents without Feynman diagrams,''
J. Phys. A \textbf{49}, no.44, 445401 (2016)
[arXiv:1510.07770 [hep-th]].



\bibitem{usprl}
R.~Gopakumar, A.~Kaviraj, K.~Sen and A.~Sinha,
``Conformal Bootstrap in Mellin Space,''
Phys. Rev. Lett. \textbf{118}, no.8, 081601 (2017)
[arXiv:1609.00572 [hep-th]].\\
R.~Gopakumar, A.~Kaviraj, K.~Sen and A.~Sinha,
``A Mellin space approach to the conformal bootstrap,''
JHEP \textbf{05}, 027 (2017)
[arXiv:1611.08407 [hep-th]].

\bibitem{usothers}
P.~Dey, A.~Kaviraj and A.~Sinha,
``Mellin space bootstrap for global symmetry,''
JHEP \textbf{07}, 019 (2017)
[arXiv:1612.05032 [hep-th]].\\
P.~Dey, K.~Ghosh and A.~Sinha,
``Simplifying large spin bootstrap in Mellin space,''
JHEP \textbf{01}, 152 (2018)
[arXiv:1709.06110 [hep-th]].

\bibitem{mack} 
G.~Mack,
``D-independent representation of Conformal Field Theories in D dimensions via transformation to auxiliary Dual Resonance Models. Scalar amplitudes,''
arXiv:0907.2407 [hep-th].

\bibitem{mellinjoao} 
J.~Penedones,
``Writing CFT correlation functions as AdS scattering amplitudes,''
JHEP {\bf 1103}, 025 (2011)
[arXiv:1011.1485 [hep-th]].

\bibitem{mellinpau} 
M.~F.~Paulos,
``Towards Feynman rules for Mellin amplitudes,''
JHEP {\bf 1110}, 074 (2011)
[arXiv:1107.1504 [hep-th]].

\bibitem{rastellizhou}
L.~Rastelli and X.~Zhou,
``Mellin amplitudes for $AdS_5\times S^5$,''
Phys. Rev. Lett. \textbf{118}, no.9, 091602 (2017)
[arXiv:1608.06624 [hep-th]].

\bibitem{alday}
L.~F.~Alday,
``On Genus-one String Amplitudes on $AdS_5 \times S^5$,''
[arXiv:1812.11783 [hep-th]].

\bibitem{RGAS}
R.~Gopakumar and A.~Sinha,
``On the Polyakov-Mellin bootstrap,''
JHEP \textbf{12}, 040 (2018)
[arXiv:1809.10975 [hep-th]].



  \bibitem{mazacpaulos}
D.~Mazac and M.~F.~Paulos,
``The analytic functional bootstrap. Part II. Natural bases for the crossing equation,''
JHEP \textbf{02}, 163 (2019)
[arXiv:1811.10646 [hep-th]].

\bibitem{mazac} D.~Maz\'a\v{c},
``A Crossing-Symmetric OPE Inversion Formula,''
JHEP \textbf{06}, 082 (2019)
[arXiv:1812.02254 [hep-th]].

\bibitem{PFKGASAZ}
P.~Ferrero, K.~Ghosh, A.~Sinha and A.~Zahed,
JHEP \textbf{07}, 170 (2020)
[arXiv:1911.12388 [hep-th]].
   
\bibitem{joaopaper}
J.~Penedones, J.~A.~Silva and A.~Zhiboedov,
``Nonperturbative Mellin Amplitudes: Existence, Properties, Applications,''
JHEP \textbf{08}, 031 (2020)
[arXiv:1912.11100 [hep-th]].
   
\bibitem{joaopaper2}
D.~Carmi, J.~Penedones, J.~A.~Silva, and A.~Zhiboedov,
``Applications of dispersive sum rules: $\epsilon$-expansion and holography,''
[arXiv:2009.13506 [hep-th]].
   
   
 \bibitem{cmrs}
S.~Caron-Huot, D.~Mazac, L.~Rastelli and D.~Simmons-Duffin,
``Dispersive CFT Sum Rules,''
[arXiv:2008.04931 [hep-th]].  

\bibitem{sleight}
C.~Sleight and M.~Taronna,
``The Unique Polyakov Blocks,''
JHEP \textbf{11}, 075 (2020)
[arXiv:1912.07998 [hep-th]].

    \bibitem{rrtv} 
  R.~Rattazzi, V.~S.~Rychkov, E.~Tonni and A.~Vichi,
  ``Bounding scalar operator dimensions in 4D CFT,''
  JHEP {\bf 0812}, 031 (2008)
  [arXiv:0807.0004 [hep-th]].


\bibitem{prv} 
  D.~Poland, S.~Rychkov and A.~Vichi,
  ``The Conformal Bootstrap: Theory, Numerical Techniques, and Applications,''
  arXiv:1805.04405 [hep-th].
   
   
   
 
   
   
   
   

\bibitem{AK}
G.~Auberson and N.~N.~Khuri,
``Rigorous parametric dispersion representation with three-channel symmetry,''
Phys. Rev. D \textbf{6}, 2953-2966 (1972)


\bibitem{ASAZ}
A.~Sinha and A.~Zahed,
``Crossing Symmetric Dispersion Relations in QFTs,''
[arXiv:2012.04877 [hep-th]].



 \bibitem{note}
In \cite{caron}, these constraints were termed as ``null constraints''. 

   
   \bibitem{TWZ}
A.~J.~Tolley, Z.~Y.~Wang and S.~Y.~Zhou,
``New positivity bounds from full crossing symmetry,''
[arXiv:2011.02400 [hep-th]].


\bibitem{caron}
S.~Caron-Huot and V.~Van Duong,
``Extremal Effective Field Theories,''
[arXiv:2011.02957 [hep-th]].

\bibitem{mellindef}
The notation in \cite{RGAS} was in terms of $s,t,u$ which are related to our $s_i$'s via
$
~s_1=s-\frac{2\Dphi}{3},~ s_2=t-\frac{2\Dphi}{3},~ s_3=u-\frac{2\Dphi}{3}.
$

\bibitem{examplenull}
 For example \cite[eq (141)]{joaopaper} is proportional  $\mathcal{M}_{-1,2}$.  In \cite{cmrs} these are called the ``odd-spin'' constraints. In fact, in the QFT context, the analogous conditions were termed as ``null constraints'' in \cite{caron}. In \cite{ASAZ}, it was shown that these are, again, precisely what we term locality constraints. Our formalism allows us to write these constraints in complete generality.
 
%
 
\bibitem{footchi}
As in \cite{ASAZ}, we find a recursion relation
\be
\chi_{n}^{(r, m)}(\tau_k)=\sum_{j=r+1}^{m}(-1)^{j+r+1} \chi_{n}^{(j, m)}(\tau_k) \frac{\mathfrak{U}^{\left(\tau_k\right)}_{n, j, r}}{\mathfrak{U}^{\left(\tau_k\right)}_{n, r, r}}
\ee




\bibitem{bissi}
A.~Bissi, P.~Dey and T.~Hansen,
``Dispersion Relation for CFT Four-Point Functions,''
JHEP \textbf{04}, 092 (2020)
[arXiv:1910.04661 [hep-th]].


   
\bibitem{Paulos:2020zxx}
M.~F.~Paulos,
``Dispersion relations and exact bounds on CFT correlators,''
[arXiv:2012.10454 [hep-th]].

\bibitem{duffin}
D.~Simmons-Duffin,
JHEP \textbf{03}, 086 (2017)
[arXiv:1612.08471 [hep-th]].


\bibitem{PHAS}
P.~Haldar and A.~Sinha,
SciPost Phys. \textbf{8}, 095 (2020)
[arXiv:1911.05974 [hep-th]].


   
  
  \bibitem{smat3}
M.~F.~Paulos, J.~Penedones, J.~Toledo, B.~C.~van Rees and P.~Vieira,
``The S-matrix bootstrap. Part III: higher dimensional amplitudes,''
JHEP \textbf{12}, 040 (2019)
[arXiv:1708.06765 [hep-th]].
\end{thebibliography}
\end{document}